\documentclass[twocolumn,showpacs,preprintnumbers,amsmath,amssymb]{revtex4-1}
\usepackage{graphicx}
\usepackage{textcomp}

\begin{document}
\title{Classical realization of two-site Fermi-Hubbard systems}
  \normalsize
\author{S. Longhi and G. Della Valle}
\address{Dipartimento di Fisica, Politecnico di Milano, Piazza L. da Vinci
32, I-20133 Milano, Italy}
\author{V. Foglietti}
\address{Istituto di Fotonica e Nanotecnologie del CNR, sezione di Roma, Via Cineto Romano 42, I-00156 Roma, Italy}

%
\bigskip
\begin{abstract}
A classical wave optics realization of the two-site Hubbard model,
describing the dynamics of interacting fermions in a double-well
potential, is proposed based on light transport in
evanescently-coupled optical waveguides.
 \noindent
\end{abstract}



\maketitle

Quantum-classical analogies have been explored on many occasions to
mimic and visualize in a purely classical setting the dynamical
aspects embodied in a wide variety of quantum systems
\cite{Dragomanbook,Longhi09LPR}. In particular, in the past two
decades engineered photonic lattices have provided a useful model
system to investigate wave optics analogous of solid state phenomena
\cite{Longhi09LPR,B01,Led1,Szameit10}. Most of the optical analogues
of solid-state phenomena observed so far, including electronic Bloch
oscillations \cite{B01,B02}, Zener tunneling \cite{ZT}, dynamic
localization \cite{DL}, Anderson localization \cite{AL}, Rabi
flopping \cite{Rabi}, and topological photonic crystals
\cite{geometric}, refer to {\em single-particle} phenomena and are
based on the formal similarity between the paraxial optical wave
equation in photonic lattices and the nonrelativistic
Schr\"{o}dinger equation of a single particle in periodic potentials
\cite{Longhi09LPR}. However, much of the richer physics in
condensed-matter comes from many-body phenomena and electron
correlations. The simplest and paradigmatic model which describes
correlation effects of electrons in a lattice, arising from the the
competition among chemical bonding, Coulomb repulsion and Pauli
exclusion principle, is perhaps provided by the Hubbard model (HM)
\cite{Hubbard}. This model is capable of capturing some many-body
aspects of the electronic properties of condensed matter, such as
metal-insulator transitions, itinerant magnetism, and electronic
superconductivity (see, e.g., \cite{L1,L2} and reference therein).
In spite of the simplicity of its Hamiltonian structure, very few
exact results are known for the HM, mainly for finite clusters or
for the infinite one-dimensional chain \cite{L1,L2,S1}. The simplest
solvable and nontrivial system, which can still capture some of the
main relevant properties of larger clusters and of the infinite
chain, is provided by the two-site Hubbard Hamiltonian (see, for
instance, \cite{Avella03}). The two-site HM, being exactly solvable,
has been considered by several authors as a simplified theoretical
model \cite{Avella03,M1,M2,M3,Kozlov96,Fox85,Ziegler10}. In
particular, it is useful as a toy model for understanding the
binding of molecules like $H_2$ \cite{M1,M2,M3}, and it was proposed
to model electron-molecular vibration coupling in organic
charge-transfer salts \cite{Kozlov96} and the electronic structure
in $\pi$ systems \cite{Fox85}. Since photons are bosons and they do
not interact when propagating in {\it linear} optical structures,
one would expect that photonics is not a suited system to simulate
in a classical setting the physics of interacting electrons in
solids. In recent works  \cite{Longhi10}, it has been pointed out
that photonic structures could provide a noteworthy laboratory
system to simulate the physics of few interacting {\em bosons} in
the framework of the Bose-Hubbard model. In this Brief Report it is
shown that light transport in suitably engineered coupled waveguide
structures can mimic the dynamics of interacting {\em fermions} as
well. In particular, an optical realization of the two-site HM is
proposed, in which light propagation in four evanescently-coupled
waveguides reproduces the temporal dynamics of the occupation number
amplitudes of the electrons in the two-site potential.
\par
The HM describing electron dynamics in a one-dimensional chain of
$N$ potential sites with nearest neighboring hopping is defined by
the Hamiltonian (see, for instance, \cite{L2})
\begin{equation}
\hat{H}=-\kappa \sum_{j=1}^{N-1} \sum_{\sigma=\uparrow ,
\downarrow}\left( \hat{a}^{\dag}_{j,\sigma} \hat{a}_{j+1,\sigma} +
\hat{a}^{\dag}_{j+1,\sigma} \hat{a}_{j,\sigma} \right)+U
\sum_{j=1}^{N} \hat{n}_{j,\uparrow}\hat{n}_{j,\downarrow}
\end{equation}
where $\kappa$ is the hopping amplitude between adjacent sites, $U$
is the on-site Coulomb interaction strength,
$\hat{a}^{\dag}_{j,\sigma}$ is the fermionic creation operator that
creates one electron at site $j$ with spin $\sigma$ ($j=1,2,...,N$,
$\sigma= \uparrow, \downarrow$), and
$\hat{n}_{j,\sigma}=\hat{a}^{\dag}_{j,\sigma}\hat{a}_{j,\sigma}$ are
the particle number operators. The fermionic operators
$\hat{a}^{\dag}_{j,\sigma}$ and $\hat{a}_{j,\sigma}$ satisfy the
usual anti-commutation relations
$\{\hat{a}^{\dag}_{j,\sigma},\hat{a}^{\dag}_{k,\rho}
\}=\{\hat{a}_{j,\sigma},\hat{a}_{k,\rho} \}=0$ and
$\{\hat{a}_{j,\sigma},\hat{a}^{\dag}_{k,\rho} \}=\delta_{j,k}
\delta_{\sigma,\rho}$. The space of states of the HM is spanned by
all linear combinations of Wannier states of the form \cite{L2}
\begin{eqnarray}
|n_1,n_2,...,n_N,m_1,m_2,...,m_N \rangle \equiv
|\mathbf{n},\mathbf{m} \rangle = \\
= \hat{a}^{\dag \; n_1}_{1,\downarrow} \hat{a}^{\dag \;
n_2}_{2,\downarrow} ... \hat{a}^{\dag \; n_N}_{N,\downarrow}
\hat{a}^{\dag \; m_1}_{1,\uparrow} \hat{a}^{\dag \;
m_2}_{2,\uparrow} ... \hat{a}^{\dag \; m_N}_{N,\uparrow} |0 \rangle
\nonumber
\end{eqnarray}

\begin{figure}
\includegraphics[scale=0.42]{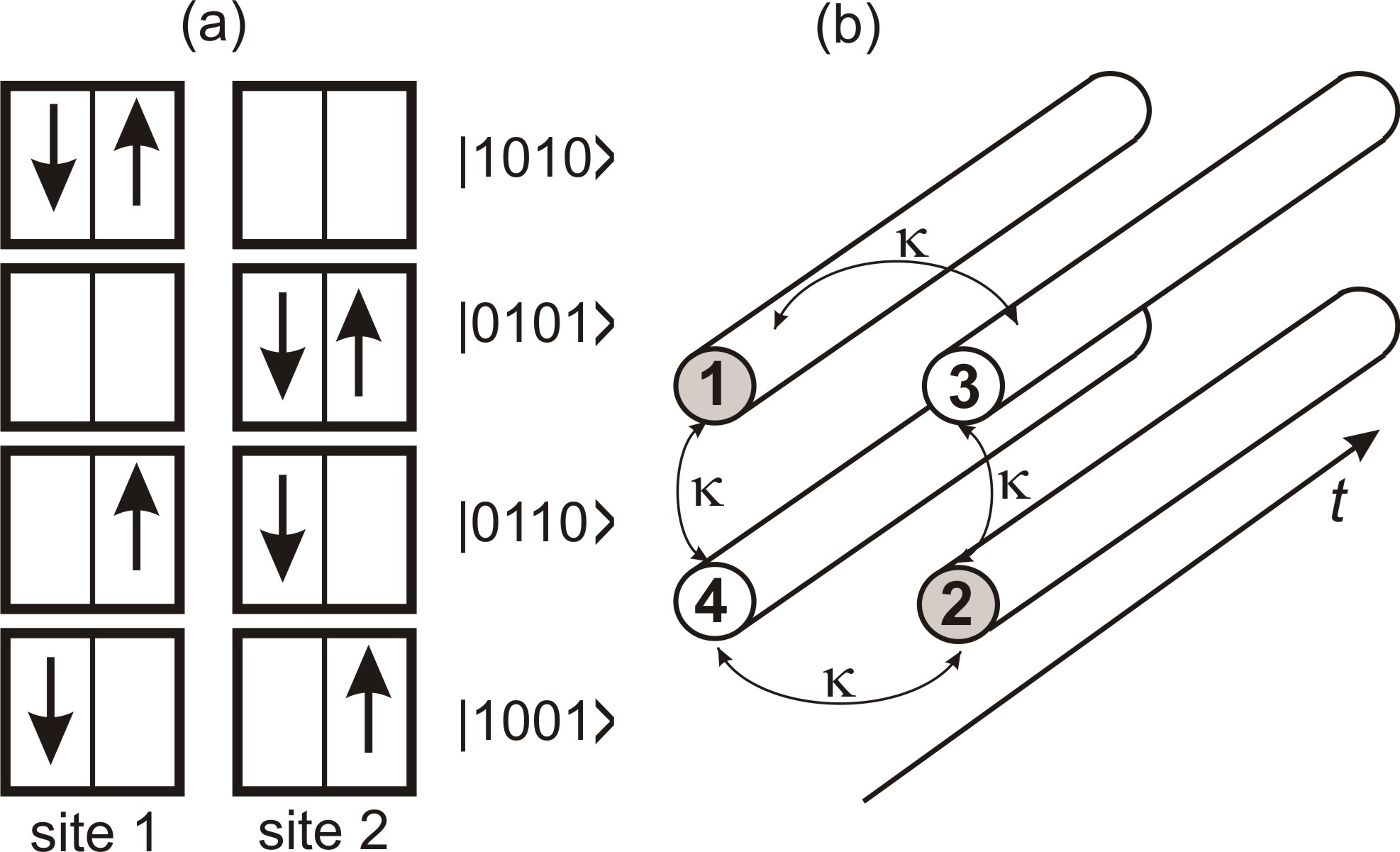}
\caption{ (a) Wannier basis of the two-site Hubbard Hamiltonian
corresponding to two electrons with opposite spins, and (b)
corresponding optical realization based on four evanescently-coupled
optical waveguides. The coupling rate between nearest waveguides is
$\kappa$, whereas the propagation constant of the modes of
waveguides 1 and 2 is shifted by $U$ from the one of waveguides 3
and 4. The waveguides 1,2,3 and 4 correspond to the four quantum
states $|1,0,1,0\rangle$, $|0,1,0,1 \rangle$, $|0,1,1,0\rangle$ and
$|1,0,0,1\rangle$, respectively.}
\end{figure}

where $|0 \rangle$ is the vacuum state. The state
$|\mathbf{n},\mathbf{m} \rangle$ corresponds to  $n_j$  electrons
occupying the site $j$ with spin $\downarrow$ and $m_j$  electrons
occupying the site $j$ with spin $\uparrow$. Owing to the
anti-commutation rules of the Fermi operators, the integers $n_j$
and $m_j$ can take only the two values $0$ and $1$, according to the
Pauli exclusion principle. Hence, the number of all linearly
independent Wannier states is $2^{2N}$. If the state vector
$|\psi(t) \rangle$ of the system is decomposed on the Wannier basis,
$|\psi(t) \rangle=
\sum_{\mathbf{n},\mathbf{m}}f(\mathbf{n},\mathbf{m},t) | \mathbf{n},
\mathbf{m}\rangle$, the evolution equations for the $2^{2N}$
occupation amplitudes $f(\mathbf{n},\mathbf{m},t)$ are formally
given by (assuming $\hbar=1$)
\begin{equation}
i \frac{df(\mathbf{n},\mathbf{m},t)}{dt}=
\sum_{\mathbf{s},\mathbf{q}} \langle
\mathbf{n},\mathbf{m}|\hat{H}|\mathbf{s},\mathbf{q} \rangle
f(\mathbf{s},\mathbf{q},t).
\end{equation}
Since the total number of electrons $N_t$ and total number of
electrons with spin $\uparrow$ ($N_{\uparrow}$) and $\downarrow$
($N_{\downarrow}$) are conserved quantities for the Hubbard
Hamiltonian \cite{note}, the amplitude equations (3) are decoupled
into a set of equations acting on different sub-spaces of the
Hilbert space. Each sub-space is defined by the Wannier states with
an assigned number of electron $N_t=N_{\downarrow}+N_{\uparrow}$,
with $N_{\uparrow}$ electrons with spin $\uparrow$ and
$N_{\downarrow}$ electrons with spin $\downarrow$; the number of
amplitudes in such a subspace
is hence $\left( \begin{array}{c} N \\
N_{\uparrow}\end{array}\right) \left(
\begin{array}{c} N \\ N_{\downarrow}\end{array}\right)$.
The two-site HM [i.e. $N=2$ in Eq.(1)] provides the simplest and
exactly solvable model which can still capture some of the main
relevant properties of larger clusters and of the infinite chain.
The two-site HM has been investigated by several authors
\cite{Avella03,M1,M2,M3,Kozlov96,Fox85,Ziegler10} and proposed as a
toy model for understanding the binding of molecules like $H_2$
\cite{M1,M2,M3}, to model electron-molecular vibration coupling in
organic charge-transfer salts \cite{Kozlov96}, and to describe the
electronic structure in $\pi$ systems \cite{Fox85}. Here we propose
an optical realization of the two-site Hubbard Hamiltonian based on
light transport in evanescently-coupled optical waveguides which is
capable of mimicking the temporal dynamics of the Hubbard system in
the Wannier basis representation (3). For the two-site Hubbard
Hamiltonian, there are 16 possible configurations of electrons on
two sites: one with no electrons, four with one electron (an up or
down electron on each of the two sites), six with two electrons (one
with two up electrons on different sites, one with two down
electrons, and four with an up electron and a down electron), four
with three electrons, and one with four electrons. The most
interesting dynamics is provided by the sub-space consisting of two
electrons with opposite spins, i.e. to $N_t=2$ and
$N_{\uparrow}=N_{\downarrow}=1$, which is spanned by the four states
$|1,0,1,0\rangle$, $|0,1,0,1 \rangle$, $|0,1,1,0\rangle$ and
$|1,0,0,1\rangle$  with amplitudes $c_1(t) \equiv f(1,0,1,0,t)$,
$c_2(t) \equiv f(0,1,0,1,t)$, $c_3(t) \equiv f(0,1,1,0,t)$ and
$c_4(t) \equiv f(1,0,0,1,t)$  [see Fig.1(a)]. In this case, the
coupled equations (3) for the amplitudes $c_l(t)$ ($l=1,2,3,4$) read
explicitly
\begin{equation}
i \frac{d}{dt} \left(
\begin{array}{c}
c_1 \\
c_2 \\
c_3 \\
c_4
\end{array}
\right)= \left(
\begin{array}{c c c c}
U & 0 & -\kappa & -\kappa \\
0 & U & -\kappa & -\kappa \\
-\kappa & -\kappa & 0 & 0 \\
-\kappa & -\kappa & 0 & 0
\end{array}
\right) \left(
\begin{array}{c}
c_1 \\
c_2 \\
c_3 \\
c_4
\end{array}
\right).
\end{equation}
\begin{figure}
\includegraphics[scale=0.46]{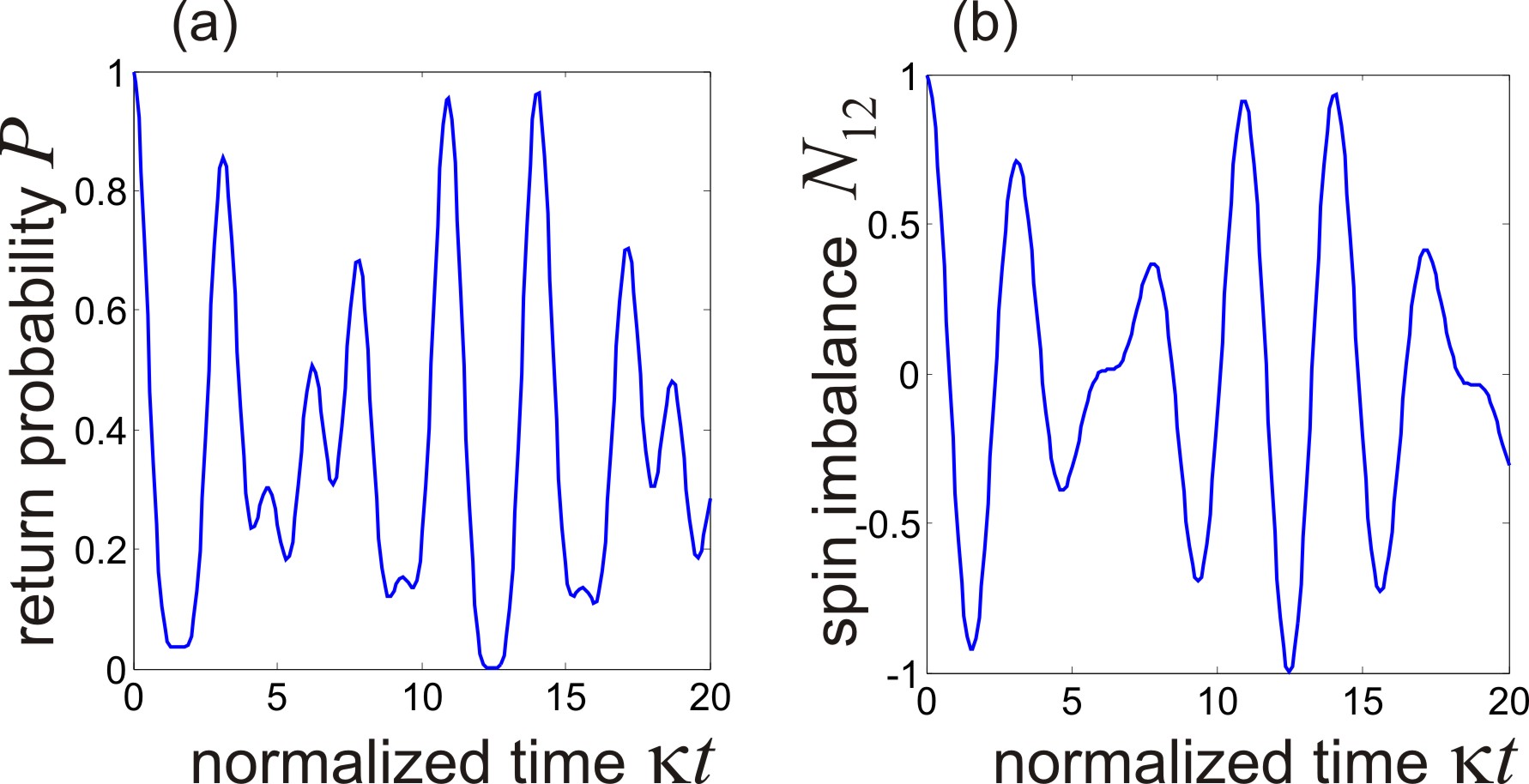}
\caption{ Temporal evolution of (a) the return probability $P(t)$,
and (b) the spin imbalance $N_{12}(t)$ for the two-site Hubbard
model for $U/\kappa=0.5$. The system is initially prepared in the
state $|0,1,1,0 \rangle$.}
\end{figure}

\begin{figure}
\includegraphics[scale=0.46]{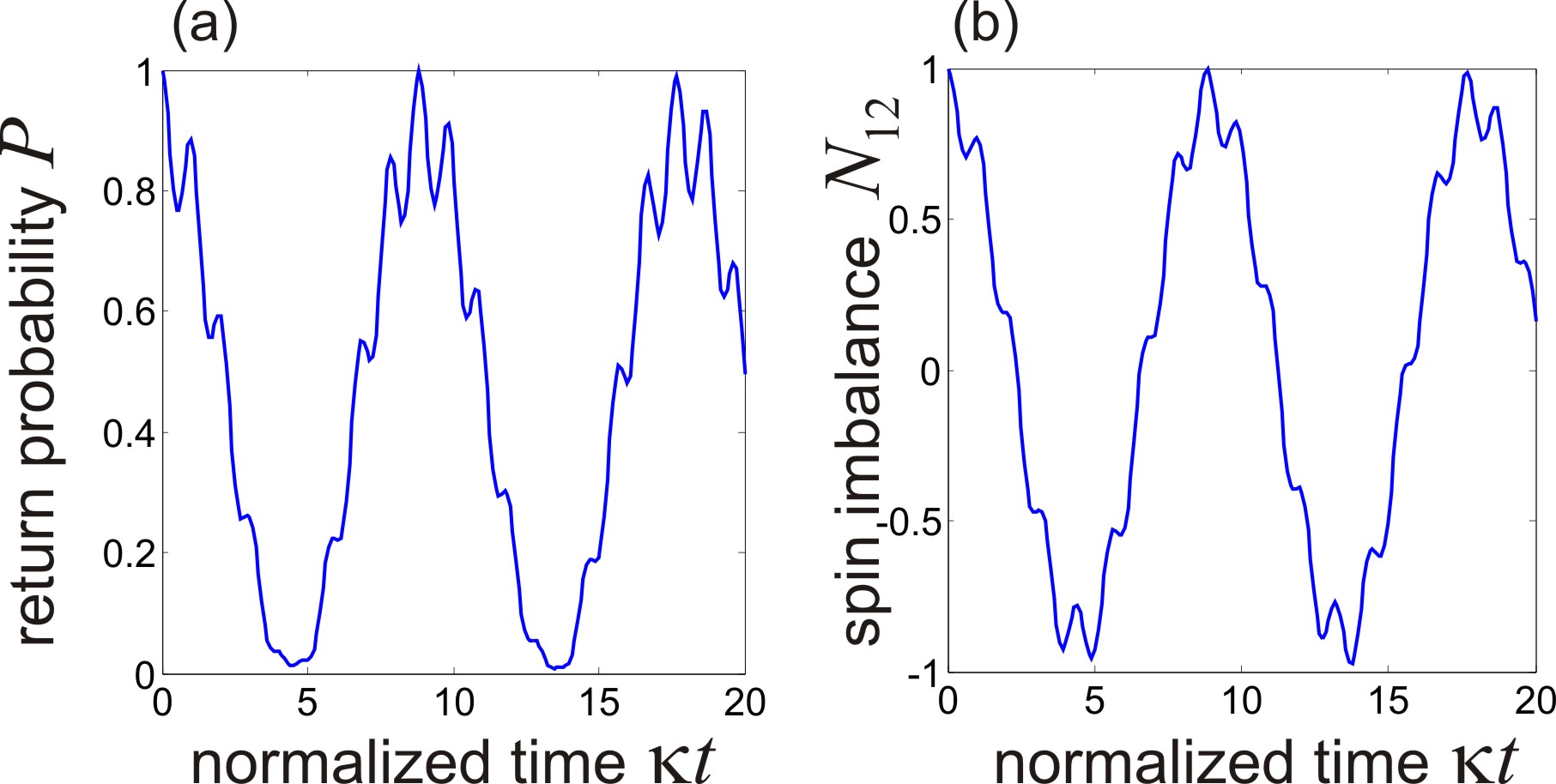}
\caption{ Same as Fig.2, but for $U/\kappa=5$.}
\end{figure}

An optical realization of the Hamiltonian system (4) is provided by
propagation of monochromatic light waves at wavelength $\lambda$ in
four evanescently-coupled optical waveguides in the geometrical
setting shown in Fig.1(b). In the optical structure, the time $t$
represents the spatial propagation distance along the waveguide
axis, whereas the Wannier amplitude $c_l$
 corresponds to the modal amplitude of light trapped in
the $l$-th waveguide ($l=1,2,3,4$). In fact, in the tight-binding
approximation the spatial part of the electric field
$\mathcal{E}(x,y,t)$ of the optical wave propagating along the $t$
spatial direction of the guiding structure can be written as
$\mathcal{E}(x,y,t) \simeq \sum_{l=1}^4 c_l(t) u_l(x,y) \exp( 2 \pi
i n_0 t/ \lambda)$, where $(x,y)$ are the spatial coordinates
transverse to the optical $t$ axis, $u_l(x,y)$ is the spatial modal
profile of the $l$-th waveguide, and $n_0$ is an effective reference
mode index. The spatial evolution of the modal amplitudes $c_l$
arising from the weak evanescent coupling of adjacent waveguides is
governed by coupled mode equations \cite{Longhi09LPR,B01,Led1} which
are formally analogous to Eqs.(4), provided that the cross-coupling
between waveguides 1 and 2, and between waveguides 3 and 4 in
Fig.1(b), is negligible. In the optical setting, the hopping rate
$\kappa$ entering in Eq.(4) is analogous to the spatial tunneling
rate of light waves between two adjacent waveguides arising from
evanescent field coupling, whereas the on-site Coulomb interaction
strength $U$ corresponds to a shift of the propagation constants of
waveguides $1$ and $2$ as compared to waveguides $3$ and $4$ [see
Fig.1(b)]. The optical structure shown of Fig.1(b) could be easily
realized in fused silica by the recently-developed femtosecond
writing technique, in which the propagation constant shift $U$ is
realized by varying the writing speed of waveguides (see, for
instance, \cite{Szameit10}). \par The energies of the two-site
Hubbard Hamiltonian can be calculated analytically as the
eigenvalues of the $4 \times 4$ matrix entering in Eq.(4), and read
explicitly (see, for instance, \cite{Kozlov96}) $E_1=0$, $E_2=U$,
$E_3=(U/2)+\sqrt{(U/2)^2+4 \kappa^2}$, and
$E_4=(U/2)-\sqrt{(U/2)^2+4 \kappa^2}$. In the optical analogy, such
energies correspond to the propagation constant mismatch of the
various supermodes of the coupled waveguides. It is worth noticing
that the energy spectrum of the simple two-site HM contains some
important physical features of more complex chains, such as the
onset of metal-insulator transition in a half-filled linear chain as
the parameter $\kappa/U$ is varied \cite{Avella03}. To see this, let
us consider the limit of small $\kappa/U$, and let us expand the
sector of Hilbert space to include all sectors with two electrons by
adding the states $|0 0 1 1\rangle$ and $|1 1 0 0 \rangle$. These
states are eigenstates of $\hat{H}$ with eigenvalue 0. All together,
the two electron space of the two site HM has four `small'
eigenvalues $0$, $0$, $0$ and $(U/2)-\sqrt{(U/2)^2+4 \kappa^2}
\simeq -4 \kappa^2/U$, and two `large' ones $U$ and
$(U/2)+\sqrt{(U/2)^2+4 \kappa^2} \simeq U$. The large eigenvalues
are associated with eigenvectors whose components have significant
mixtures of the states with doubly occupied sites. The existence of
the two groups of states whose eigenvalues are separated by $U$ is a
reflection of the upper and lower Hubbard bands in a lattice. The
'Mott-Hubbard' gap in the spectrum gives rise to a metal-insulator
transition \cite{L2}. For a given initial state $|\psi(t=0)
\rangle=|\psi_0 \rangle$, there are two interesting observables
related to the dynamical evolution of the two-site HM, namely the
return probability $P(t)=|\langle \psi(t)| \psi_0 \rangle|^2$ and
the spin imbalance between the two sites, $N_{12}(t) = (1/2) \langle
\psi(t)|
\hat{n}_{1,\uparrow}-\hat{n}_{1,\downarrow}+\hat{n}_{2,\downarrow}-\hat{n}_{2,\uparrow}
|\psi(t) \rangle$ (see, for instance, \cite{Ziegler10}). The latter
describes the exchange dynamics of the two spins $\uparrow$ and
$\downarrow$, located at the two sites. As an example, Figs.2 and 3
show a typical behavior of the return probability $P(t)$ and spin
imbalance $N_{12}(t)$ for the two-site HM corresponding to a weak
($U/\kappa=0.5$, Fig.2) and a strong ($U/\kappa=5$, Fig.3) on-site
Coulomb interaction for a system initially prepared in the Wannier
state $|0,1,1,0 \rangle$, i.e. for the initial condition
$c_l(0)=\delta_{l,3}$. In the Wannier basis representation, the
return probability $P(t)$ and spin imbalance $N_{12}(t)$ take the
simple form
\begin{equation}
P(t)=\sum_{l=1}^4 |c_l^*(0)c_l(t)|^2 \; ,\;
N_{12}(t)=|c_3(t)|^2-|c_4(t)|^2.
\end{equation}
Note that in the optical analogue the spin imbalance has a very
simple meaning: it is just the power imbalance of light between
waveguides 3 and 4. Moreover, if the system is initially prepared in
one of the Wannier state, the return probability is simply mapped
into the fractional optical power trapped in the initially excited
waveguide. For example, if the system is initially prepared in the
state $|0,1,1,0 \rangle$ as in Figs.2 and 3, one has
$P(t)=|c_3(t)|^2$. It is worth noticing that, the optical analogue
of the strong on-site Coulomb interaction regime $U/\kappa \gg 1$
(as in Fig.3) corresponds to a nearly-sinusoidal exchange of optical
power between waveguides 3 and 4, like in an ordinary synchronous
optical direction coupler \cite{Yariv}. In fact, in the large
$U/\kappa$ limit and for the initial condition $c_1(0)=c_2(0)=0$,
the amplitudes $c_1$ and $c_2$ remain small and can be eliminated
from the dynamics by standard perturbation methods. This yields the
reduced dynamical equations for the amplitudes $c_3$ and $c_4$
\begin{eqnarray}
i \frac{dc_3}{dt}  & \simeq -\kappa_{e}c_4 +\delta c_3\\
i \frac{dc_4}{dt}  & \simeq -\kappa_{e} c_3+ \delta c_4
\end{eqnarray}
where $\kappa_{e} \equiv 2 \kappa^2/U$ is an effective coupling
constant and $\delta=-\kappa_{e}$ a common propagation constant
detuning. Hence light coupling in the four-waveguide structure in
the large on-site Coulomb interaction regime is like the one of an
ordinary two-waveguide directional coupler, showing Rabi-like
exchange of the optical power between the (not directly coupled)
waveguides 3 and 4 mediated by the two off-resonance waveguides 1
and 2. The observed oscillatory power exchange between two uncoupled
waveguides, arising from indirect coupling via weakly excited
off-resonance waveguides, is analogous to the process of two-photon
Rabi oscillations observed in Ref.\cite{Ornigotti}.
\par

In conclusion, a simple classical realization of the two-site
Fermi-Hubbard Hamiltonian, based on light propagation in
evanescently-coupled optical waveguides, has been theoretically
proposed. While previous theoretical and experimental studies on
optical simulations of quantum phenomena in solid-state physics have
been concerned with single-particle phenomena, here it has been
shown that photonics can provide a laboratory tool to visualize and
simulate in simple optical settings the dynamical aspects embodied
in the physics of interacting fermionic systems. The present study
has been focused on a simple Hubbard system, however it is envisaged
that photonic simulators of others and more complex models of
interacting fermions can be realized. Possible extensions of the
present study include the photonic realization of the
Hubbard-Holstein Hamiltonian \cite{HH}, which describes bipolarons
dynamics arising from electron-phonon coupling, and the
Hubbard-Anderson Hamiltonian \cite{Moura} describing the dynamics of
two electrons on a linear chain with long-range correlated disorder
and on-site Hubbard interaction. The Hubbard-Anderson Hamiltonian
can be simulated using a two-dimensional square array of waveguides
with engineered propagation constants, which can be used to test in
an optical setting the interplay between disorder, localization and
electron-electron interaction. Phonon-electron coupling dynamics for
a two-site Hubbard-Holstein Hamiltonian can be realized by coupling
two of the four waveguides of Fig.1(b) with two semi-infinite linear
arrays with non-homogeneous hopping rates, which simulate the
vibrational (phonon) degrees of freedoms of the two sites.
\par

 The authors acknowledge financial support by the
Italian MIUR (Grant No. PRIN-2008-YCAAK project "Analogie
ottico-quantistiche in strutture fotoniche a guida d'onda").


\begin{thebibliography}{}
\bibitem{note} This is because the electrons are conserved in number and
cannot change their spin when they hop.
\end{thebibliography}
\end{document}